\documentstyle[prl,aps,epsfig]{revtex}

\begin{document}

\draft

\title{The survival of quantum coherence in deformed states 
superposition}

\author{
Stefano Mancini$^1$ 
and Vladimir I. Man'ko$^2$}

\address{
${}^1$INFM, Dipartimento di Fisica,
Universit\`a di Milano,
Via Celoria 16, I-20133 Milano, Italy
\\
${}^2$P. N. Lebedev Physical Institute, Leninskii Prospekt 53,
Moscow 117924, Russia}

\date{\today}

\maketitle

\widetext

\begin{abstract}
We study the dissipative dynamics of deformed coherent states 
superposition. We find that such kind of superposition can be        
more robust against decoherence than the usual Schr\"odinger
cat states.
\end{abstract}

\pacs{42.50.Dv, 42.50.Ar, 03.65.Ca}

\maketitle
\widetext

The feature of quantum mechanics which most distinguishes it from 
classical mechanics is the coherent superposition of distinct 
physical states. Many of the less intuitive aspects of the quantum 
theory can be traced to this feature. 
The famous Schr\"odinger cat argument \cite{SCH}
highlights problems of interpretation 
where macroscopic superposition states is allowed. 
In fact, such states are very fragile in the presence of 
dissipation, and rapidly collapse to a classical mixture exhibiting 
no unusual interference features \cite{CL,WALLS}.
Environment induced decoherence has been identified in recent years 
as one of the main ingredients in the transition from quantum to 
classical behavior \cite{GIU,ZUR}. Classicality emerges as a 
consequence of the coupling 
of quantum systems to an environment which, in effect, dynamically 
enforces super-selection rules by precluding the stable existence 
of the majority of states in the Hilbert space of the system. 
The physics of decoherence has been studied during the last few 
years both from the theoretical \cite{ZUR} 
and also from the experimental point 
of view \cite{BRUNE}.
However, the most used paradigmatic case has been the 
superposition of two distinguishable coherent states
\cite{DMM}. The latter being eigenstates of boson 
annihilation operator \cite{GLA}.

On the other hand, quantum groups \cite{DRIN}, introduced as a 
mathematical description of deformed Lie algebras, have given the 
possibility of generalizing the notion of creation and annihilation 
operators of the usual oscillator and to introduce deformed 
oscillator \cite{BIE}. 
They were interpreted \cite{NAPLES} as nonlinear oscillators 
with a very 
specific type of nonlinearity, and this led to the more general
concept of $f$-deformed oscillator \cite{MMZS}.
Then, the notion of $f$-coherent states was straightforwardly 
introduced \cite{MMZS}, and the generation of such nonlinear 
coherent states enters
in the real possibilities of trapped systems \cite{VOGEL}. 

Successively, it was quite natural to consider the superposition 
of such states \cite{PLA}, which 
could be named {\it deformed cat state}. 
Here, we are going to study their dissipative dynamics and 
the related coherence properties in comparison with the 
usual Schr\"odinger cat states \cite{DMM}.
In particular, we shall show
that the deformed cat states may result much more robust 
against decoherence than their undeformed version.

The essential point in understanding quantum coherence is 
the physical
distinction between the coherent superposition state 
\begin{equation}\label{RHOPURE}
|\psi\rangle=\sum_i c_i|\psi_i\rangle
\Longleftrightarrow
\rho=\sum_{i,j}c_ic^*_j |\psi_i\rangle\langle\psi_j|
\,,
\end{equation}
and the classical mixture 
\begin{equation}\label{RHOMIX}
\rho_{mix}=\sum_i|c_i|^2 |\psi_i\rangle\langle\psi_i|\,.
\end{equation}
Following the avenue sketched in Ref.\cite{WALLS},
the density operator (\ref{RHOPURE}) can be written as 
\begin{equation}\label{RHOSPLIT}
\rho=\rho_{mix}+\sum_{i\neq j}c_i c_j^* 
|\psi_i\rangle\langle\psi_j|\,.
\end{equation}
Let $Z$ be the operator corresponding to some
physical observable having  
eigenvalues $z$. Then, the probability distribution for 
$Z$ in the state
$|\psi\rangle$ is given by
\begin{equation}\label{PZ}
P(z)=P_{mix}(z)+ 
\sum_{i\neq j}c_i c_j^* \langle z|\psi_i\rangle
\langle\psi_j|z\rangle\,,
\end{equation}
where $P_{mix}(z)=\sum_i|c_i|^2|\langle z|\psi_i\rangle|^2$.
Measurement of $Z$ will distinguish the states $\rho$ and 
$\rho_{mix}$
provided the second term in (\ref{PZ}) is nonzero.
We are thus led to define the quantum coherence function 
(with respect to the measurement $Z$) as 
\begin{equation}\label{CZ}
{\cal C}(z)=\sum_{i\neq j}c_i c_j^* \langle z|\psi_i\rangle
\langle\psi_j|z\rangle\,.
\end{equation}

The usual coherent state is defined \cite{GLA}
as the eigenstate of the
annihilation operator $a$, of a bosonic field 
\begin{equation}\label{COH}
|\alpha\rangle=N\sum_{n=0}^{\infty}\frac{\alpha^n}{\sqrt{n!}}
|n\rangle\,,
\quad
N=\left[ \exp(\alpha^2)\right]^{-1/2}\,,
\end{equation}
where we have assumed 
$\alpha\in{\bf R}$ for the sake of simplicity.
Then, it is possible to consider
the superposition
\begin{equation}\label{CAT}
|\Psi\rangle=N_+\left(|\alpha\rangle
+|-\alpha\rangle\right)\,,
\quad
N_+=\left[2+2N^2 \exp({-\alpha^2})\right]^{-1/2}\,,
\end{equation}
known as {\it even cat state} Ref.\cite{DMM}. 
This is a remarkable example of interference among 
quantum states.

Now, consider to measure the number operator,
i.e. $Z\equiv a^{\dag}a$, then, 
the number probability distribution results
\begin{equation}\label{PN}
P(n)=P_+(n)+P_-(n)+2{\rm Re}\left\{{\cal C}(n)\right\}\,,
\end{equation}
with $P_{\pm}(n)=|\langle n|\pm\alpha\rangle|^2$ 
and ${\cal C}(n)=\langle \alpha|n\rangle\langle n|-\alpha\rangle$.
Hence, a convenient measure of quantum coherence is the 
quantum visibility 
\begin{equation}\label{VDEF}
{\cal V}=\frac{|{\cal C}(n)|}{\sqrt{P_+(n)P_-(n)}}\,.
\end{equation}
A straightforward calculation gives
${\cal V}=1$, which means that the state (\ref{CAT}) 
shows maximum quantum coherence.

In order to apply the same arguments at macroscopic level, 
it is also useful to define a parameter giving the separation 
among the two states being superposed. 
However, the concept of ``distance" between quantum states 
is not uniquely defined 
(a recent discussion on this 
problem can be found in Ref.\cite{DOD}),
and here, we adopt the simplest one, leading to
\begin{equation}\label{DIST}
d=\langle\alpha|\left(a+a^{\dag}\right)|\alpha\rangle
=2\alpha\,.
\end{equation}
It is worth noting that such distance is directly related to 
the number of photons $\alpha^2$ of the coherent states.

We now introduce the decoherence effects
due to a dissipative interaction with an environment \cite{ZUR}.
This can be described (in interaction picture)
by the following master equation of the Lindblad form \cite{GAR}
\begin{equation}\label{ME}
\dot{\rho}=\gamma\, a\rho a^{\dag}
-\frac{\gamma}{2} \left\{a^{\dag}a,\rho\right\}\,,
\end{equation}
where $\gamma$ is the damping rate,
and we have set the bath temperature equal to zero.
The decoherence effect on the state
$\rho(0)=|\Psi\rangle\langle\Psi|$
can be described in the following way \cite{GJM}
\begin{equation}\label{RHOT}
\rho(t)=\sum_{k=0}^{\infty}
\Upsilon_k(t)\rho(0)\Upsilon^{\dag}_k(t)\,,
\end{equation}
where
\begin{equation}\label{UPS}
\Upsilon_k(t)=\sum_{n=k}^{\infty}
\sqrt{
\left(
\begin{array}{c}
n\\k
\end{array}
\right)}\;
\left[\eta(t)\right]^{(n-k)/2}\,\left[1-\eta(t)\right]^{k/2} \,  
|n-k\rangle\langle n|\,,
\end{equation}
with $\eta(t)=e^{-\gamma t}$.

Thus, the quantum visibility results
\begin{equation}\label{V}
{\cal V}(n,t)=\exp\left\{-2\alpha^2\left[1-\eta(t)\right]\right\}\,.
\end{equation}
This is a well known results \cite{WALLS,PHOENIX} 
showing that the decoherence effect
depends on the damping rate as well as on the separation
of the coherent states, 
i.e. the macroscopicity. Moreover, ${\cal V}$ remains 
constant through $n$.
Considering that $\gamma$ is fixed by the system-environment 
interaction, we would investigate whether the quantum visibility
can depend on the type of cat state one consider.

Let us introduce a $f$-coherent state 
defined \cite{MMZS} as the eigenstate of
the annihilation operator of a $f$-deformed
bosonic field $A=a\sqrt{f(a^{\dag}a)}$,
where $f$ is an operator-valued function of the number operator
(here it is assumed Hermitian and real).
In general, it can be made dependent on continuos parameters, 
in such a way that, for given particular values, the usual algebra 
is recovered.  
The $f$-coherent state can be written as  
\begin{equation}\label{FCOH}
|\zeta, f\rangle={\cal N}\sum_{n=0}^{\infty}
\frac{\zeta^n}{\sqrt{[n]_f!}}|n\rangle\,,
\quad
{\cal N}=\left[\exp_f(\zeta^2)\right]^{-1/2}\,,
\end{equation}
where we have considered $\zeta\in {\bf R}$, and we have introduced
\begin{equation}\label{EXPF}
\exp_f(x)\equiv\sum_{n=0}^{\infty}\frac{x^n}{[n]_f!}\,,
\quad\quad
[n]_f!\equiv\left[nf(n)\right]\times
\left[(n-1)f(n-1)\right]\times\ldots
\left[2 \, f(2)\right]
\times\left[f(1)\right]
\times\left[f(0)\right]\,.
\end{equation}
The function $\exp_{f}$ is a deformed version of 
the usual exponential function.
They become coincident when $f$ is the identity.
Notice that $\exp_f(x)\exp_f(y)\ne \exp_f(x+y)$,
i.e. we have a non-extensive exponential 
which can be found in many physical problems
\cite{TSALLIS}.

Let us now consider the superposition
\begin{equation}\label{FCAT}
|\Phi\rangle={\cal N}_+\left(|\zeta,f\rangle
+|-\zeta,f\rangle\right)\,,
\quad
{\cal N}_+=\left[2+2{\cal N}^2
\exp_f(-\zeta^2)
\right]^{-1/2}\,.
\end{equation}
In this case the separation between the two superposed states
becomes
\begin{equation}\label{FDIST}
d=\langle\zeta,f|\left(a+a^{\dag}\right)|\zeta,f\rangle
={\cal N}^2 \, 
\sum_{n=0}^{\infty}\frac{\zeta^n}{\sqrt{[n]_f!}}
\left\{
\frac{\sqrt{n}\,\zeta^{(n-1)}}{\sqrt{[n-1]_f!}} 
+\frac{\sqrt{n+1}\,\zeta^{(n+1)}}{\sqrt{[n+1]_f!}}
\right\}\,.
\end{equation}

The decoherence effects on the state 
$\rho(0)=|\Phi\rangle\langle\Phi|$
are introduced by again employing
Eq.(\ref{RHOT}).
Here, we are not interested on the 
dynamics in presence of a deformed 
Hamiltonian \cite{SCRIPTA}.
Then, the quantum visibility 
can be obtained as straightforward extension of
previous argument
\begin{equation}\label{VF}
{\cal V}(n,t)=
\left|\sum_{k=0}^{\infty}
\frac{(n+k)!}{k!}
\frac{\left[-\zeta^{2}\left(1-\eta(t)\right)\right]^k}
{[n+k]_{f}!}\right|
\times
\left\{\sum_{k=0}^{\infty}
\frac{(n+k)!}{k!}
\frac{\left[\zeta^{2}\left(1-\eta(t)\right)\right]^k}
{[n+k]_{f}!}\right\}^{-1}\,.
\end{equation}
From the above equation, it is clear that the 
quantum visibility depends on
the outcome of the measurement of the observable $Z$.
Moreover ${\cal V}$, depending on the specific form
of $f$, could be greater than the non-deformed 
case. It follows the possibility to preserve the
quantum coherence.
Of course, in order to correctly compare 
the two situations, i.e. deformed and undeformed, 
one should consider cat states 
of the same macroscopicity,
i.e. having the same $d$.

Then, let us consider two types of deformations in more details.
The $q$-deformation defined by \cite{BIE,NAPLES}
\begin{equation}\label{FQ}
f(n)=\sqrt{\frac{1}{n}\frac{q^n-q^{-n}}{q-q^{-1}}}\,,
\quad
f(0)=1\,,
\quad
q\in {\bf R}\,,
\end{equation}
and the deformation given by \cite{VOGEL},
\begin{equation}\label{FLAG}
f(n)=\frac{L^{1}_{n}(\xi^{2})}{(n+1)L^{0}_{n}(\xi^{2})}\,,
\quad 
\xi\in{\bf R}\,,
\end{equation}
which we are going to name $L$-deformation,
since $L^m_n$ indicates the associate 
Laguerre polynomial. 
It is worth noting that such $L$-deformation naturally arises
in ion trapped systems \cite{VOGEL}.

\begin{figure}[t]
\centerline{\epsfig{figure=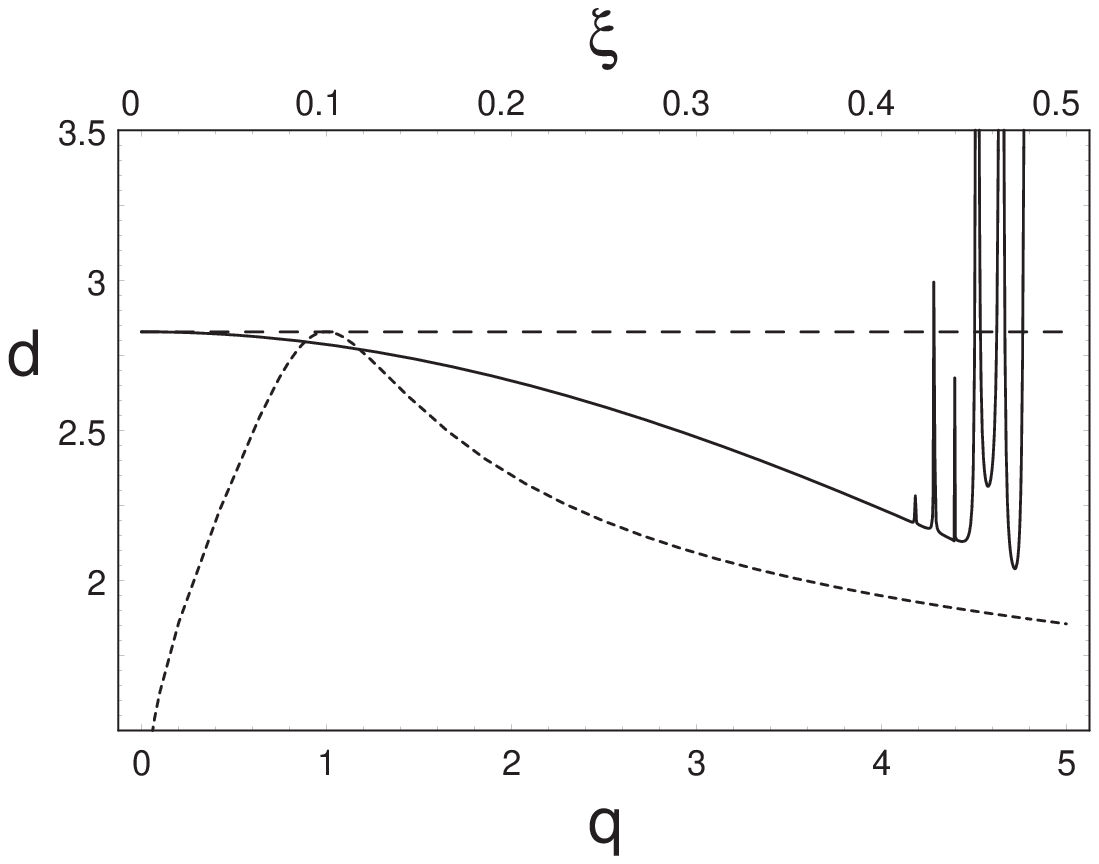,width=3.5in}}
\caption{\widetext 
States separation versus parameter deformation.
The dashed line represents the non-deformed case; the dotted 
line the $q$-deformation; the solid line the $L$-deformation.
$\zeta^2=\alpha^2=2$.
}
\label{fig1}
\end{figure}

From Eq.(\ref{V}) we can argue a survival of coherence
by decreasing $\zeta$ with respect to $\alpha$
still maintaining the same states separation $d$.
In Fig.(\ref{fig1}) we show the separation 
as function of deformation parameter when
$\zeta^2=\alpha^2=2$.
The dashed line represents the value of undeformed cat state.
Then, we may see that the $q$-deformation (dotted line)  
always implies a smaller distance $d$,
while the $L$-deformation (solid line) allows to reach,
and overcome, the distance of the undeformed 
coherent states ($2\sqrt{2}$). 
This is due to the form of Eq.(\ref{FLAG}) which
shows many singularities by varying $\xi$.
Thus, for our purpose, the $L$-deformation seems more pertinent
since allows a comparison between deformed and undeformed cat 
states having the same separation.

\begin{figure}[t]
\centerline{\epsfig{figure=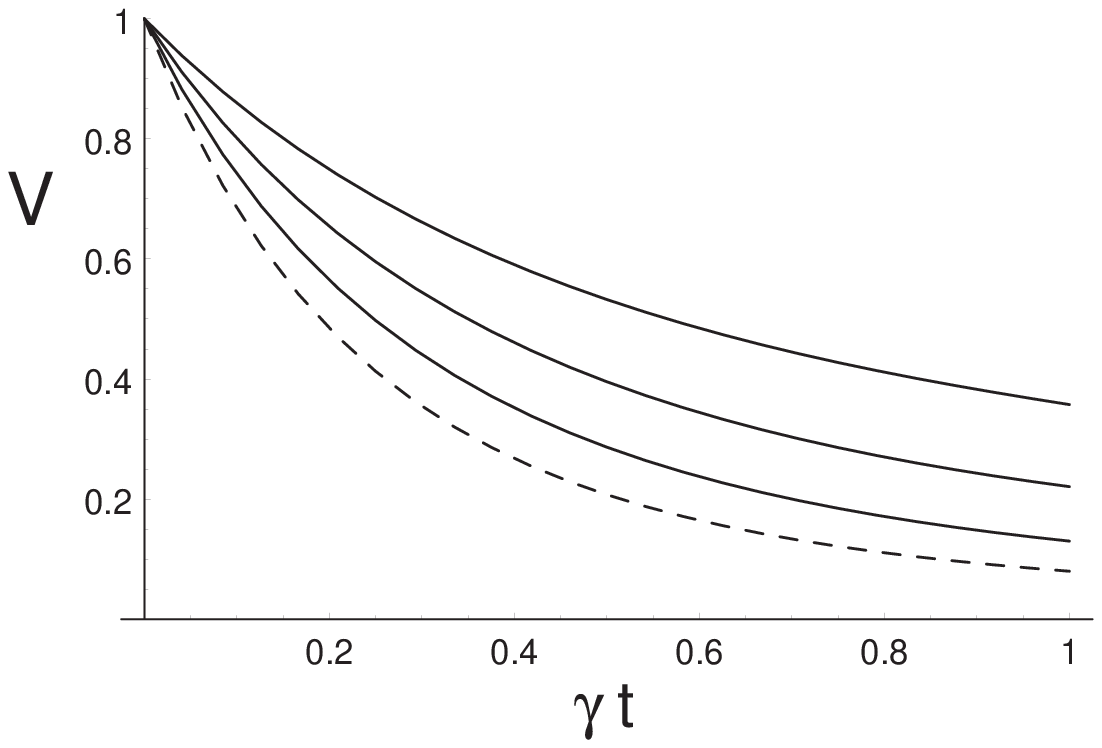,width=3.5in}}
\caption{\widetext 
Quantum visibility as function of dimensionless time $\gamma t$.
The dashed line represents the non-deformed case; the solid lines
the $L$-deformed case with $\zeta^2=\alpha^2=2$ and $\xi=0.45048$.
From top to bottom solid lines refer to $n=3$, $n=2$ and $n=1$.
}
\label{fig2}
\end{figure}

In Fig.(\ref{fig2}) we show the quantum visibility as function of
dimensionless time $\gamma t$. We immediately see that the 
decoherence of $L$-deformed cat state (solid lines) is slowed down 
with respect to the undeformed one (dashed line). Moreover,
in the deformed case the quantum visibility depends on $n$,
and it is better preserved by increasing $n$.
Of course the relevant value is $n=2$ (since it is the most probable
for both case) and it shows a consistent improvement.

\begin{figure}[t]
\centerline{\epsfig{figure=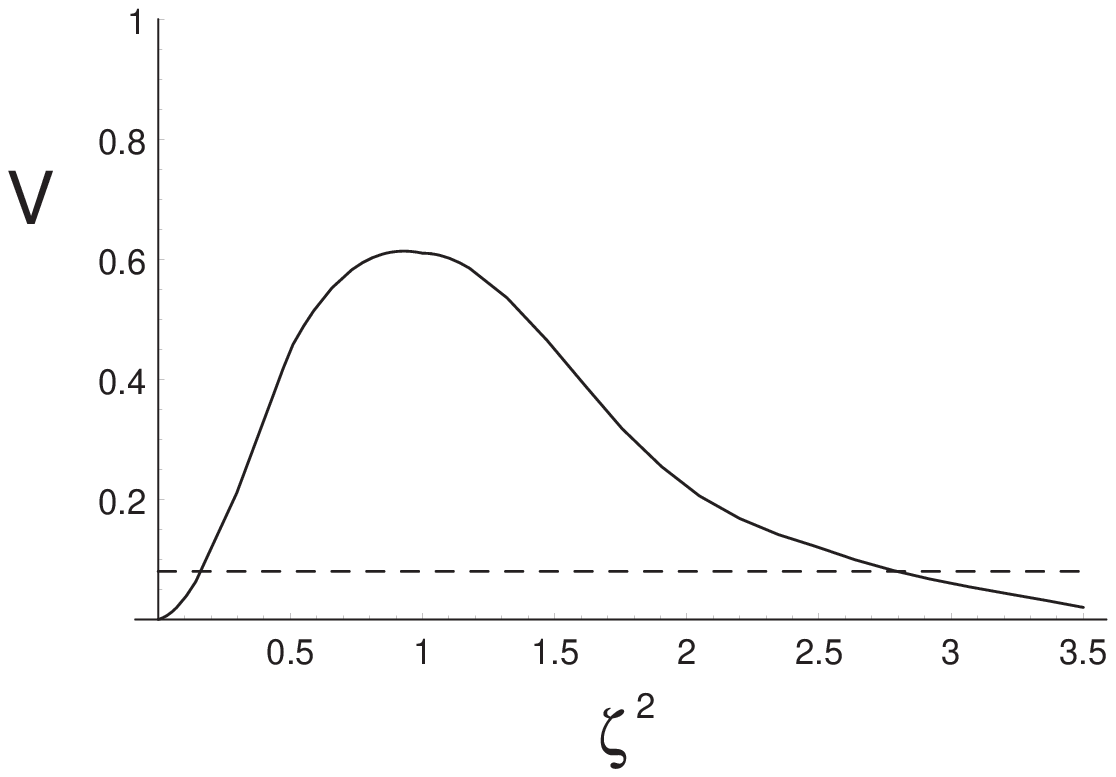,width=3.5in}}
\caption{\widetext 
Quantum visibility at $\gamma t=1$ as function of $\zeta^2$.
The dashed line represents the non-deformed case with $\alpha^2=2$; 
the solid lines refers to $n=2$, for $L$-deformed case. 
}
\label{fig3}
\end{figure}

Nevertheless, as stated above, we expect a better result for 
$\zeta<\alpha$. Then, in Fig.(\ref{fig3}), we have plotted
the quantum visibility at $\gamma t=1$ as function of $\zeta$.
For each value of $\zeta$ we have used the value of $\xi$ 
giving the same distance $d$ of the undeformed case
($2\sqrt{2}$).
We may seen that there exist a value of $\zeta^2$ which maximize
the coherence persistence. For such a value ($\zeta^2\approx 1$), 
the visibility results enormous greater than the undeformed case.
For $\zeta^2\to 0$ the visibility tends to zero 
since practically the two superposing states 
become the vacuum states.
Instead, for $\zeta^2 \gg 2$, the deformed 
case has always a distance
$d$ greater than the deformed one, so that no comparison is 
possible.

In conclusion we have shown that quantum coherence can survive
against dissipation provided to superpose distinguishable 
coherent states of suitable deformed field.
This unexpected result relies on the fact that the
states being superposed, once deformed,
are no longer eigenstates (nor near eigenstates)
of the operator appearing in the irreversible 
part of the evolution  equation (\ref{ME}),
i.e. they substantially differ from the pointer basis
\cite{ZUR}.
On the other hand, deformed states, due to their nonlinear
character, give rise to a more rich phase space structure 
\cite{VOGEL}, part of which can easier survive against 
decoherence.

The present results may open new perspectives 
for the experimental observation 
of macroscopic realism in quantum mechanics.
Moreover, the kind of studied states, being decoherence resistant, 
could result quite useful for 
quantum information processing \cite{INFO}.
Extension of the above arguments to other observables, or
to other types of decoherence, e.g. non-dissipative 
decoherence \cite{BONI}, is planned for future work.

\end{document}